\documentstyle[aps,preprint,epsfig,eqsecnum]{revtex}
\begin{document}
\draft
\title{Nucleon-nucleon scattering observables from solitary boson
exchange potential}
\author{L. J\"ade and H. V. von Geramb}
\address{Theoretische Kernphysik, Universit\"at Hamburg\\
Luruper Chaussee 149, D-22761 Hamburg, Germany}
\date{\today}
\maketitle
\begin{center}
Published in Phys.\,Rev.\ C {\bf 57}, 496 (1998)\\
\end{center}
\begin{abstract}
The one solitary boson exchange potential (OSBEP) is used
to evaluate observables of {\it NN} elastic scattering
below pion threshold. In this approach, we use
a nonlinear model of self-interacting mesons
as a substitution for the commonly used
phenomenological form factors. {\it NN} data
support an empirical scaling law between the pion and other meson
fields, which suggests a link to QCD and 
significantly reduces the number of parameters in the 
boson exchange potential. 
The analysis of $np$ and $pp$ observables distinguishes
the model by its fit and few adjustable
parameters. An outlook to apply OSBEP in $\pi N$ systems is
given. 
\end{abstract}
\narrowtext
\section{Introduction}
The notion of interacting elementary particles for low- and
medium-energy nuclear physics is associated with
definitions of potential operators which, inserted into a
Lippmann-Schwinger equation, yield the scattering phase shifts and
observables. In principle, this potential carries the rich
QCD substructure consisting of quarks and gluons and thus may
be deduced from some microscopic model. There are a number of
models which explicitly refer to QCD and have
gained remarkable success describing qualitative features of
hadronic interactions \cite{ch}. 
Unfortunately, so far none of these models
is able to reach the accuracy of phenomenological boson
exchange or inversion potentials \cite{pot,inv}. These models,  
however, do not contain any explicit reference to QCD and in the case of
boson exchange models use 
effective baryon and meson fields with phenomenological masses,
coupling constants and form factors. 
It remains astonishing that, with these assumptions, they 
are able to account for a highly quantitative
description of {\it NN} data below pion production threshold 
and thus have established themselves as
the standard models to be used in nuclear
physics. Furthermore, inversion and boson exchange models work 
equally well for meson-nucleon \cite{inv2,mnp} and meson-meson 
\cite{inv2,mmp} interactions. 
This implies that the potentials remain valid 
at relative distances of $\sim$0.3\,fm, which is much
smaller than the rms radii of mesons and nucleons themselves
and smaller than the QCD bag sizes. It is beyond any doubt
that nucleons and mesons are genuine QCD objects and we
expect their effects to become distinguishable within relative
distances of $\sim$1.5\,fm. In this context, it is common belief that
phenomenological form factors effectively
describe the actual QCD dynamics at short distances. 
  
To perform a step towards QCD inspired models,
we attempt to replace the conventional form factors by a
nonlinear meson dynamics using the one solitary boson 
exchange potential (OSBEP), which
was developed recently by the Hamburg group \cite{Jae97}. 
From the success of the empirical boson exchange
potentials, it seems obvious that chiral symmetry is not dominant in 
{\it NN} scattering below 300\,MeV \cite{hhi}. 
None the less, also a phenomenological
low-energy model should be inspired by concepts 
which ensure chiral symmetry conservation.
In this sense, we adopt structures from the linear 
$\sigma$ model and develop a dynamics of 
self-interacting mesons. At this stage, we cannot circumvent 
chiral symmetry breaking by 
taking the nonlinearities, masses, and coupling constants as free
parameters.
Most important, the self-interaction is taken into account persistently 
at all instances. This is achieved by using meson fields
which are quasiclassical analytic solutions of 
nonlinear field equations. Defining free meson operators, 
the quantization of these fields is done {\it a posteriori}.
Finally, this model is utilized in the 
framework of a one boson exchange potential (OBEP),
which closely follows the Bonn-B potential \cite{Mach89}.  
  
The benefit of this approach is the inclusion of
nonlinear effects, leading to meson propagators of finite self-energy, 
which permits us to replace 
the form factors in conventional boson exchange potentials. 
Furthermore, an empirical scaling law was discovered 
which relates the pion mass and its self-interaction
coupling constant with the self-interaction parameters of
any of the other mesons used. Confirming our conjecture about
reminiscent effects of the microscopic substructure
subsumed in the empirical form factors, we interpret this 
as a hint for an underlying symmetry. An obvious
benefit of the scaling law is the practical bisection of
the number of adjustable parameters. This is a 
different approach than pursued
by the Bonn-CD \cite{bcd} or Reidlike Nijmegen potentials
\cite{nipot}, which achieve perfect fits with an inflating number of
parameters.     
   
A description of the theoretical framework, including
technical details, can be found in \cite{Jae97}. 
Therein, we restricted the
analysis to fit $np$ SM95 phase shifts only.
In the present work, we extend the potential to describe $np$ as
well as $pp$ scattering and calculate scattering
observables to be compared with the latest database compiled in {\sc SAID}
\cite{said}. Additionally, we show  $np$ and $pp$ phase shift
comparisons for Bonn-B \cite{Mach89}, Nijm93
\cite{nipot}, Paris \cite{papot}, OSBEP, and the
analysis SM97 of Arndt {\it et al.} \cite{sm97}. 
        
This paper is organized as follows. 
In Sec.\,\ref{sm}, we give the salient features 
of OSBEP. The fit of the model parameters to phase shifts
is discussed in Sec.\,\ref{nnph} and thereafter, an
extensive survey of $np$ and $pp$ scattering observables
is given in Sec.\,\ref{obs}. An outlook for 
application of the nonlinear model to $\pi N$ scattering, 
together with a summary, is contained in Sec.\,\ref{outl}.
\section{Solitary Mesons}
\label{sm}
It is a common feature of chiral invariant models that
spontaneous symmetry breaking leads to nonlinear terms in
the mesonic part of a Lagrangian which can be interpreted as a
self-interaction \cite{Itzz80}. Because of this, a meson
Lagrangian with the same structure as the linear
$\sigma$ model for all mesons in the OBE potential is
assumed. Altogether, we consider the six mesons
$\beta=\pi,\eta,\rho,\omega,\sigma,\delta$, and a   
Lagrangian 
\begin{equation}
{\cal L}_{\beta} = \frac{1}{2}\left(\partial_{\mu}\Phi_{\beta}
\partial^{\mu}\Phi_{\beta}-m_{\beta}^2\Phi_{\beta}^2\right)
-\frac{\lambda^{\beta}_1}{2p+2}\Phi_{\beta}^{2p+2}-
\frac{\lambda^{\beta}_2}{4p+2}
\Phi_{\beta}^{4p+2}+{\cal L}_{\mbox{\scriptsize int}}.
\label{ml}
\end{equation}
For mesons with nonzero spin the operator $\Phi_{\beta}$ 
is a vector in Minkowski space. 
The parameter $p$ assumes {\small 1/2} or 1 to distinguish 
odd and even powered nonlinearities and 
${\cal L}_{\mbox{\scriptsize int}}$ contains desirable couplings to
nucleon and other meson fields. 
In chiral symmetric models, the
self-interaction coupling constants $\lambda^{\beta}_1$ and
$\lambda^{\beta}_2$ and the various meson masses 
are related by symmetry relations.
This sounds intriguing but is not practical. In view of the
ambiguities contained in Eq.\ (\ref{ml}) and, in particular in ${\cal
L}_{\mbox{\scriptsize int}}$, 
it appears wise to restrict oneself first to a 
quantitative model which allows chiral symmetry breaking.
In actual calculations, this implies 
the permission of free parameters in Eq.\ (\ref{ml})
which are the coupling constants, physical
masses, and the nonlinearities $\lambda^{\beta}_i$.
After fitting the parameters to observables, 
we rely on their nature to effectively restore chiral 
symmetry {\it a posteriori} \cite{mmp}.  
\subsection{Meson propagation}
The Lagrangian (\ref{ml}) contains self-interacting
mesons and possible couplings between themselves and to
nucleons. Following the standard one boson exchange
models, we neglect meson-meson correlations and treat the
interaction between mesons and nucleons perturbatively. 
The self-interaction of each meson makes the difference
to standard models as it is taken into account in a closed
analytic form and persistently, leading to analytical solutions of the
nonlinear field equations for each Fourier component of the meson fields
in Eq.\ (\ref{ml}). For the explicit form and the quantization of these
solutions, dubbed as {\it solitary meson fields}, we refer to our 
former publication \cite{Jae97}. 
       
The probability for the propagation of solitary mesons 
can now be defined as the amplitude to create an interacting field 
at some space-time point $x$ which is annihilated into the
vacuum at $y$. The momentum space amplitudes of the  
solitary meson propagator then reads 
\begin{equation}
\label{pprop}
iP_{\beta}(k^2,m_{\beta})=\sum_{n=0}^{\infty}
\Big[C_n^{1/{2p}}(w_{\beta})\Big]^2
\frac{\left[(m_{\beta}^p\alpha_1^{\beta})^2
-m_{\beta}^{2p}\alpha_2^{\beta}\right]^n 
(2pn+1)^{2pn-2}}{{D_{k,n}^{(\beta)}}^{2pn+1}(\vec{k}\,^2
+M_{n,\beta}^2)^{pn}}\;i\Delta_F(k^2,M_{n,\beta}),
\end{equation}
where we introduced the dimensionless coupling constants
\begin{eqnarray}
\alpha_{\beta} & = & \displaystyle{\frac{1}{(2m_{\beta}V)^p}
\sqrt{\left(\displaystyle{\frac{\lambda^{\beta}_1}
{4(p+1)m_{\beta}^2}}\right)^2
-\displaystyle{\frac{\lambda^{\beta}_2}
{4(2p+1)m_{\beta}^2}}}}, 
\nonumber\\
 & & \nonumber\\
\alpha_1^{\beta} & = & \frac{\lambda^{\beta}_1}
 {4(p+1)m_{\beta}^2(2m_{\beta}V)^p}, \nonumber\\
 & & \nonumber\\
\alpha_2^{\beta} & = & \frac{\lambda^{\beta}_2}
{4(2p+1)m_{\beta}^2(2m_{\beta}V)^{2p}},
\label{a1a2}
\end{eqnarray}
and
\begin{equation}
\label{wa}
w_{\beta}=\frac{\alpha_1^{\beta}}{\sqrt{{\alpha_1^{\beta}}^2-
\alpha_2^{\beta}}}.
\end{equation}
The Feynman propagator 
\begin{equation}
\label{fprop}
i\Delta_F(k^2,M_{n,\beta})=\frac{i}{k^2-M_{n,\beta}^2},
\end{equation}
uses the mass spectrum
$$
M_{n,\beta}=(2pn+1)m_{\beta}.
$$
For $p=1/2$ one gets the amplitude for 
scalar fields and $p=1$ describes pseudoscalar particles. Vector
mesons require $p=1$ and each term of the sum is
multiplied with the Minkowski tensor 
$$
\left(-g^{\mu\nu}+
\frac{k^{\mu}k^{\nu}}{M_{n,v}^2}\right).
$$
The series (\ref{pprop}) converges rapidly and  
in practical calculations it is sufficient to use 
$n\leq 4$.

The Lorentz invariant normalization $D_{k,n}^{(\beta)}$, 
which occurs in the propagator (\ref{pprop}), is obtained from the 
normalization $D_k^{(\beta)}$ of the solitary meson fields by 
substituting
$$
k^{\mu}\quad\to\quad\frac{1}{2pn+1}k^{\mu}.
$$
  
At this point, we simplify our model. The linear
$\sigma$-model implies that the nonlinear
term associated with $\lambda_2^{\beta}$ in Eq.\ (\ref{ml}) is
zero for all mesons despite the scalar $\sigma$ and
$\delta$ mesons. Since the former is an effective particle to
simulate two-pion exchange and the latter contributes
little, it is no disadvantage to put
$\lambda_2^{\beta}=0$, implying $\alpha_2^{\beta}=0$,
for all mesons used. 
This restriction simplifies our
expressions and we notice 
$\alpha_1^{\beta}=\alpha_{\beta}$ which allows to 
drop the subscripts from
$\lambda_1^{\beta}$ and $\alpha_1^{\beta}$. 
\subsection{Proper normalization}
\label{pn}
The momentum dependent normalization $D_k^{(\beta)}$ of the solitary
mesons plays an important role and requires a detailed discussion. 
$D_k^{(\beta)}$ can depend on the four-momentum $k^{\mu}$ and the
coupling constant $\alpha_{\beta}$. The following
conditions \cite{Burt81} are imposed: 
(i) all amplitudes are to be Lorentz invariant,
(ii) $D_k^{(\beta)}$ is dimensionless,
(iii) all Feynman-diagrams are to be finite, and 
(iv) the fields are vanishing for $\alpha_{\beta}\to 0$. 
   
The amplitude (\ref{pprop}) has to fulfill on-shell
conditions known from renormalization theory
\cite{Itzz80}. At 
$k^2=m_{\beta}^2$, the propagator $iP(k^2,m_{\beta})$ has
to have a pole with residue $i$. Defining
\begin{equation}
\Gamma_{\beta}^{(2)}(k^2)=i\left[P_{\beta}(k^2,m_{\beta})\right]^{-1},
\end{equation}
these conditions can be met using
\begin{equation}
\label{bed1}
\Gamma_{\beta}^{(2)}(k^2)\Bigg|_{k^2=m_{\beta}^2}  =  0
\end{equation}
and
\begin{equation}
\frac{d}{dk^2}\Gamma_{\beta}^{(2)}(k^2)\Bigg|_{k^2=m_{\beta}^2}  =  1.
\end{equation}
From Eq.\ (\ref{pprop}) it is clear, that $iP(k^2,m_{\beta})$ readily
fulfills Eq.\ (\ref{bed1}).
The second condition demands that $D_k^{(\beta)}$
equals one for $k^2=m_{\beta}^2$.
  
Furthermore, to obtain finite results for all self-energy
diagrams involving solitary mesons, it is sufficient to choose
$$
D_k^{(\beta)}={\cal O}(k^2),
$$
for spinless particles, and
$$
D_k^{(\beta)}={\cal O}(k^4),
$$
for vector mesons. In summary, all conditions are met with
\begin{equation}
D_k^{(\beta)} = \Bigg\{1+
\left(\displaystyle{\frac{1}{\alpha_{\beta}4(p+1)(2m_{\beta})^p}}\right)
^{2/p}\left(\sqrt{\vec{k}\,^2+m_{\beta}^2}-k_0\right)^2
\Bigg\}^{S+1},
\label{dk}
\end{equation}
where $S$ denotes the particle spin.
With this proper normalization, the solitary meson propagator is
completely determined and can be applied in a boson exchange
potential. 
\subsection{The scaling law}
\label{ben}
In conventional models, meson exchange is described by 
a product of a Feynman
propagator and an empirical form factor
\begin{equation}
\label{ff}
\frac{i}{k^2-m_{\beta}^2}
\left(\frac{\Lambda_{\beta}^2-
m_{\beta}^2}{\Lambda_{\beta}^2+\vec{k}\,^2}\right)^{2n_{\beta}}.
\end{equation}
Using the proper normalization (\ref{dk}), the solitary meson
propagator was found to resemble very closely the expression (\ref{ff})
used in the Bonn-B potential \cite{Mach89}. This essential result 
permitted us to drop the phenomenological 
form factors. The astonishing benefit unfolds when we make this 
comparison for all mesons. 
Doing so, one sees an empirical
scaling relation for the self-interaction coupling
constants \cite{Jae97}
\begin{equation}
\label{scaling}
\alpha_{\beta} = 
\alpha_{\pi}\sqrt{S+1}
\left(\displaystyle{\frac{m_{\pi}}{m_{\beta}}}\right)^p.
\end{equation}
$\alpha_{\pi}$ is the only remaining parameter to 
describe the full meson dynamics. This
reduces significantly the number of parameters with respect
to Bonn-B potential.
\section{$NN$ phase shifts}
\label{nnph}
In the calculation of $NN$ phase shifts, we use the meson masses 
of the Bonn-B potential. There is some evidence that the $\pi NN$ 
coupling constant should have a value below the previously used 
one $g_{\pi}^2/{4\pi}=14.4$. The first indication came from
a Nijmegen analysis \cite{nigpi} which suggests $f_{\pi NN}^2=0.0745$ 
(which yields $g_{\pi}^2/{4\pi}=13.79$ with our values for the 
pion and nucleon mass). Additionally, Arndt and 
co-workers deduced similar value 
from their analysis of $\pi N$ scattering \cite{gpi}. 
Since we confirmed their result in an independent analysis 
\cite{SAN97} and 
intend to apply OSBEP in $\pi N$ interactions, we fix the   
$\pi NN$ coupling constant to the Arndt value
\begin{equation}
\label{pinn}
\frac{g_{\pi}^2}{4\pi} = 13.75.
\end{equation}
The parameter $\alpha_{\pi}$ and the remaining  
meson-nucleon coupling constants then yield a total number of eight 
adjustable parameters.

As in the first analysis \cite{Jae97}, 
we started our fitting procedure with $np$ phase
shifts \cite{sm97} and deuteron properties, disregarding $pp$ data. 
In this case, we are free of
Coulomb effects and there are more partial waves due to
the isoscalar and isovector contributions.  
In this work, the $pp$ data were also considered. 
This required to replace (i) the average nucleon mass
938.926\,MeV by the proton mass 938.272\,MeV, (ii) the
average pion mass 138.03\,MeV by the $\pi^0$ mass 134.98\,MeV. 
Additionally, the $\sigma_1$ coupling constant was reduced
by 1.3\,\% from its $np$ value. 
A static point charge Coulomb potential 
was included using the
Vincent-Phatak method to calculate the Coulomb distorted 
hadronic phase shifts \cite{VP74}.

\begin{figure}[t]\centering
\begin{picture}(16.0,20.0)(0.0,2.0)
\epsfig{figure=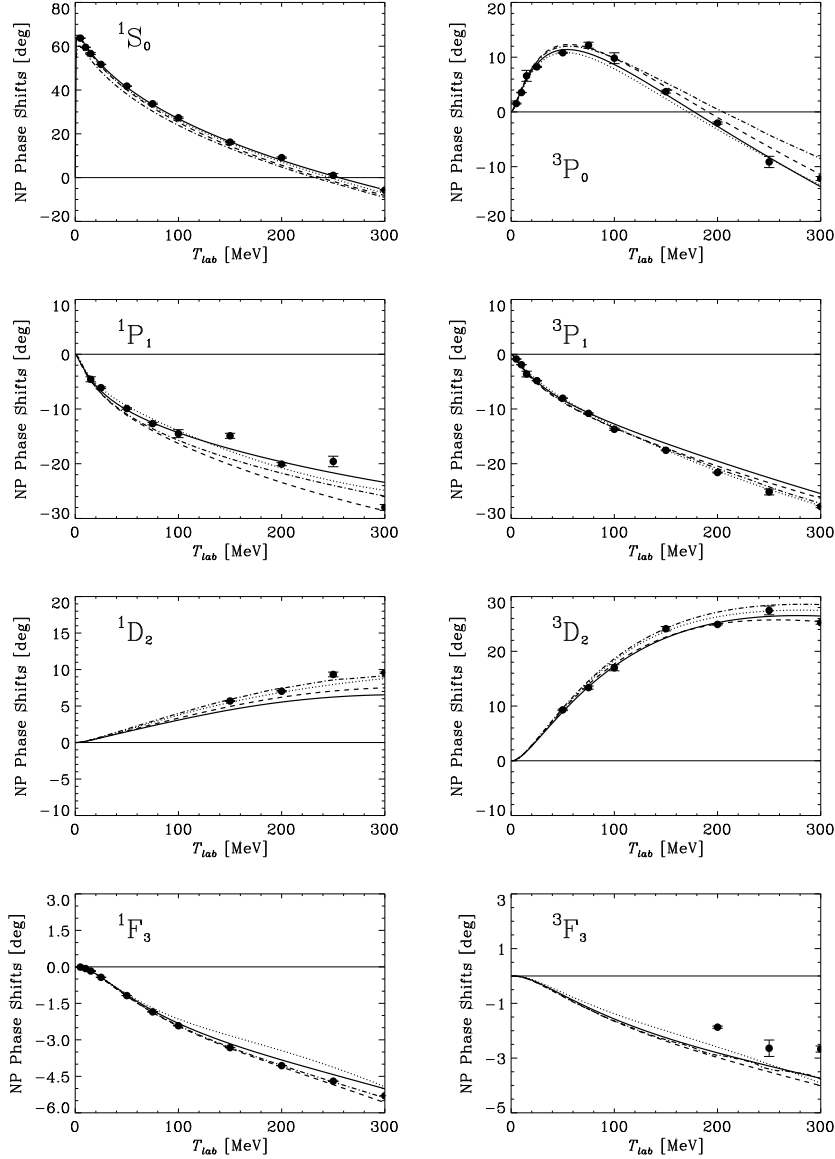,width=16.0cm}
\end{picture}
\caption[$np$ Phase Shifts]{$np$ phase 
shifts. We show the Arndt SM97 \cite{said} phase shift
analysis (circles) compared to the potentials  Nijm93 
(dotted), Bonn-B (dashed), Paris (dash-dotted),  
and OSBEP (full).} 
\label{npph}
\end{figure}

\begin{figure} \centering
\begin{picture}(16.0,20.0)(0.0,2.0)
\epsfig{figure=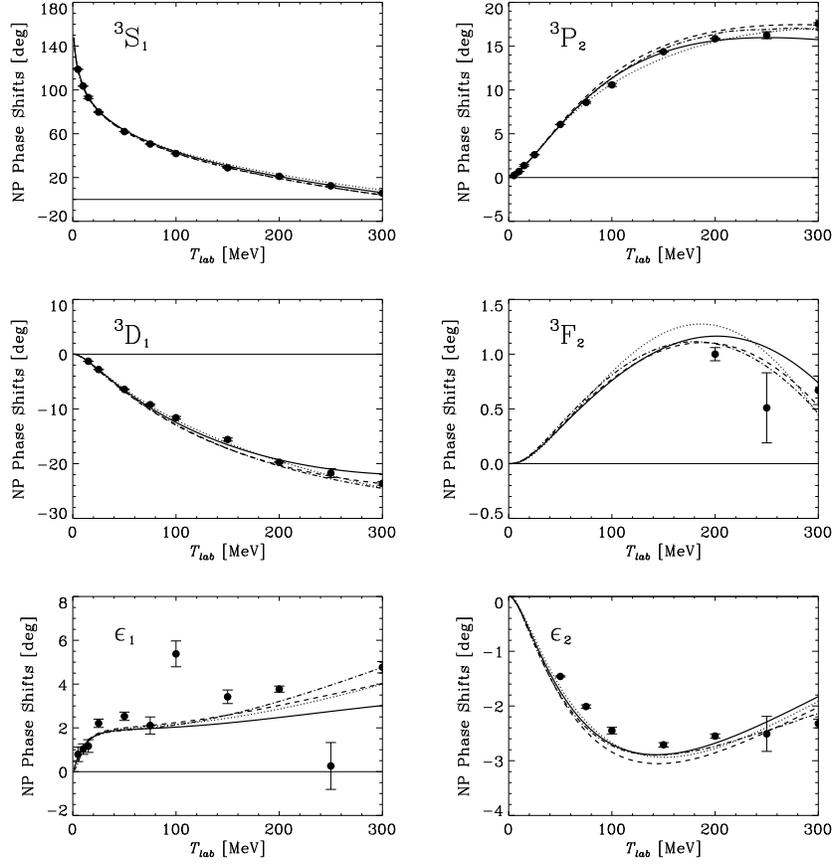,width=16.0cm}
\end{picture}
\caption{SYM $np$ phase shifts for the coupled $^3SD_1$ and 
$^3PF_2$ channels, notations as in Fig.\,\protect\ref{npph}.
\label{coupl}}
\end{figure}

\begin{figure}\centering
\begin{picture}(16.0,20.0)(0.0,2.0)
\epsfig{figure=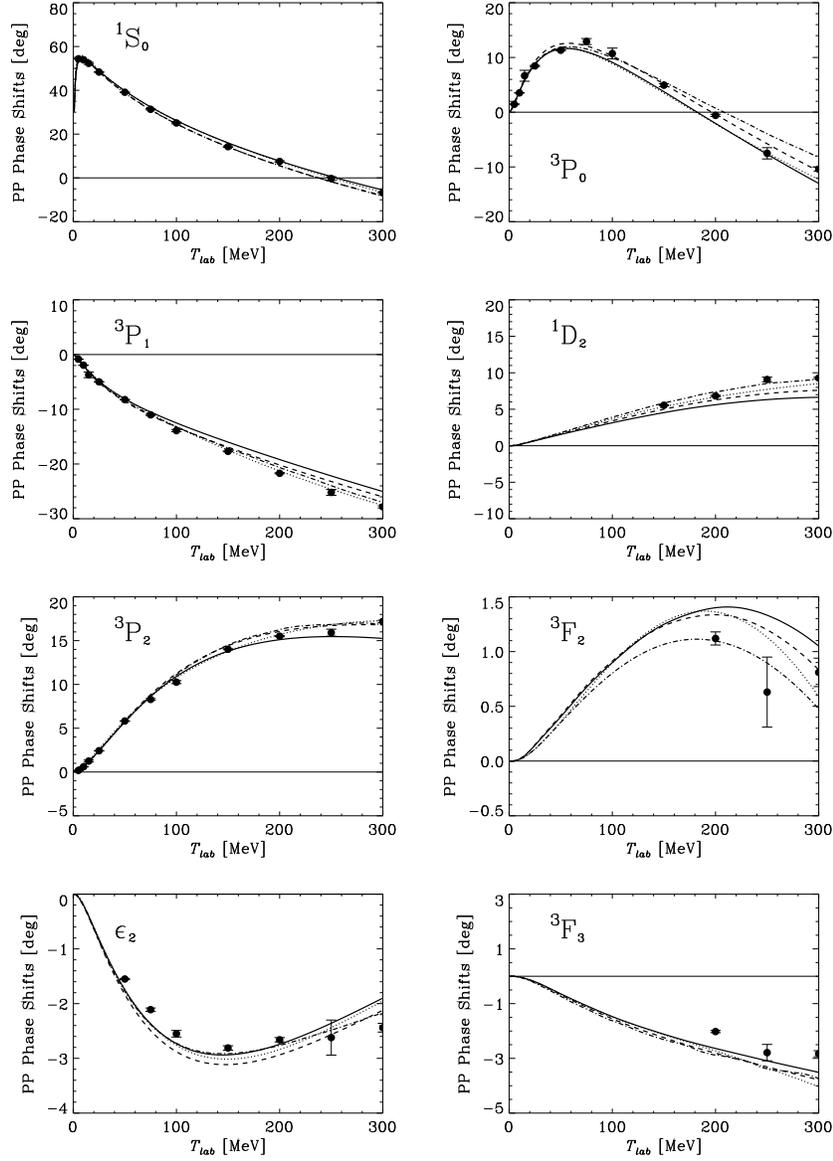,width=16.0cm}
\end{picture}
\caption{$pp$ phase shifts, notations as in
Fig.\,\protect\ref{npph}.\label{ppph}}
\end{figure}

The final parameter set is listed in
Table\,\ref{coco}. Deuteron properties are very well
reproduced and are given in Table\,\ref{deut}. 
$np$ phase shifts are shown in
Figs.\,\ref{npph} and \ref{coupl} for single and
coupled channels, respectively. $pp$ phases are contained in
Fig.\,\ref{ppph}. We plot the results from 
the Bonn-B, Nijm93, Paris, and OSBEP potential as
well as the single energy SM97 analysis. 
All potentials are in close agreement. 
Differences, with values of several degrees, 
do exist for the $^1S_0$ phase shifts 
of which the Paris potential is the worst.  
This is true over the whole energy range. Experimentally, the  
change of sign for $np$ lies
at $T_{\mbox{\scriptsize lab}}=255.2$\,MeV 
\cite{said}. At 250\,MeV, the theoretical 
values are -2.47 (Paris), -1.72 (Bonn-B), -0.45 (Nijm93), 0.73
(OSBEP), and $1.03\pm 0.84$ (SM97). 
Most striking are the deviations in the $P$ channels.
They become crucial at energies above 50\,MeV and are visible in
the observables. There exist a large amount of literature about  
these deviations, but a convincing and final solution has not been put
forward. In particular, it is known that a
potential without $\pi\rho$ correlations leads to an
overattraction in the $P$ waves \cite{Mach87}. 
This can be expected to have
a major effect in $pp$ scattering, since isoscalar $D$ waves
are absent. It is surprising that, 
despite the large and consistent database which determines the
phase shifts and which is well described by the potential models,
there are none the less some strong deviations within the model phase
shifts. 
    
Our fitting procedure leaves the coupling constants in
qualitative agreement with most Bonn-B values
\cite{Mach89}. Differences occur for the 
$\pi NN$ and $\eta NN$ coupling constants and 
the tensor to vector ratio $\kappa$. 
We use the experimental value $g_{\pi}^2/{4\pi}=13.75$, whereas
Bonn-B uses 14.4 which was preferred in the 1980's. 
Differential cross section data at backward angles 
support the lower value. SU(3) flavor symmetry 
relates the $\pi NN$ and $\eta NN$ coupling constants to be
\begin{equation}
\label{etann}
\frac{g_{\eta}^2}{4\pi}=\frac{1}{3}(3-4\alpha_f)^2
\frac{g_{\pi}^2}{4\pi},
\end{equation}
with $\alpha_f \sim$0.6-0.65. This yields 
$0.7 \leq g_{\eta}^2/{4\pi}\leq
1.7$, consistent with the value in Table \ref{coco}. 
More support for the small value can be found in literature
\cite{etann}. Increasing the $\eta NN$ coupling serves
to simulate $\pi\rho$ contributions which are generally absent in 
one boson exchange potentials \cite{Mach87}.
In the Bonn-B potential, the value $g_{\eta}^2/{4\pi}=3$ is used.
       
Another 
feature of our parameter set is the low tensor to vector ratio
$\kappa=3.8$ which is in close agreement with 
the vector-dominance value
3.7, to be compared with $\kappa=6.1$ in Bonn-B. This 
is reconciled by introducing a
direct vector coupling of the photon to the nucleon \cite{Mach87}.
We agree with the Nijmegen group that a small $\pi NN$ coupling constant
should be aligned with a value of $\kappa$ close to the vector dominance
value \cite{nigpi}. 
\section{Observables of $NN$ scattering}
\label{obs}
To obtain observables from phase shifts we follow the notation 
of Hoshizaki \cite{Hos68}. The program {\sc SAID} \cite{said} contains
explicitly this option but offers additionally the convention of
Bystricky {\it et al.} \cite{byst}. Experimental data with error bars 
and normalizations together with the theoretical phase
shifts for Nijm93 and Paris were taken from {\sc SAID}. Bonn-B
we calculated ourselves and verified its agreement with
published values \cite{Mach89}. 
\subsection{$np$ observables}
\label{npob}
\begin{figure}[t] \centering
\begin{picture}(16.0,20.0)(0.0,2.0)
\epsfig{figure=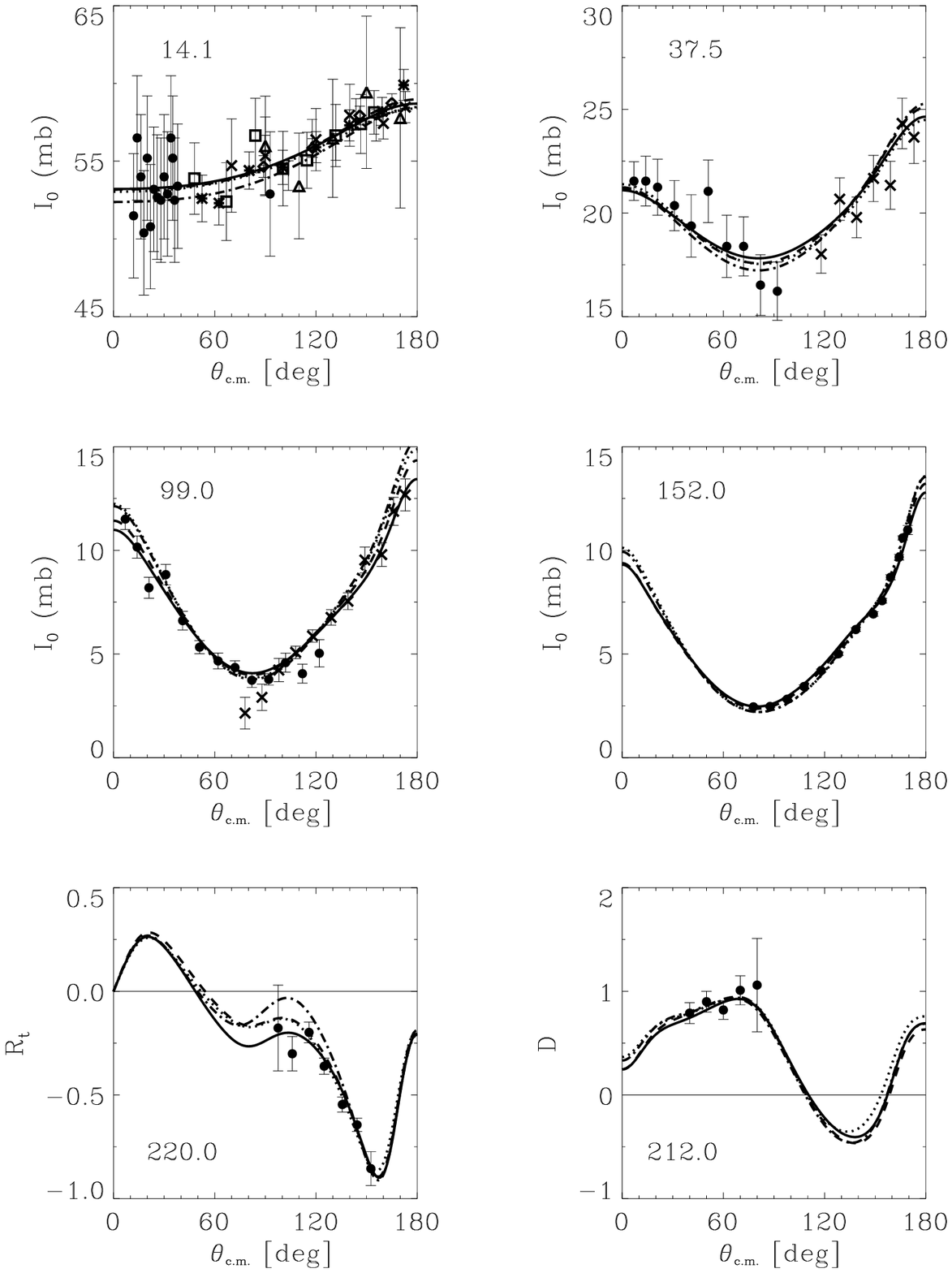,width=16.0cm}
\end{picture}
\caption[Observables of $np$ scattering]
{Observables of $np$ scattering. 
Kinetic laboratory energy is denoted, experimental data are taken from
\protect{\cite{said}} with notation from \protect\cite{Hos68}. 
We show theoretical predictions from OSBEP
(full) and Bonn-B (dashed), Nijm93 (dotted), and Paris
(dashed-dotted). \label{npob1}}
\end{figure}

\begin{figure}[t] \centering
\begin{picture}(16.0,20.0)(0.0,2.0)
\epsfig{figure=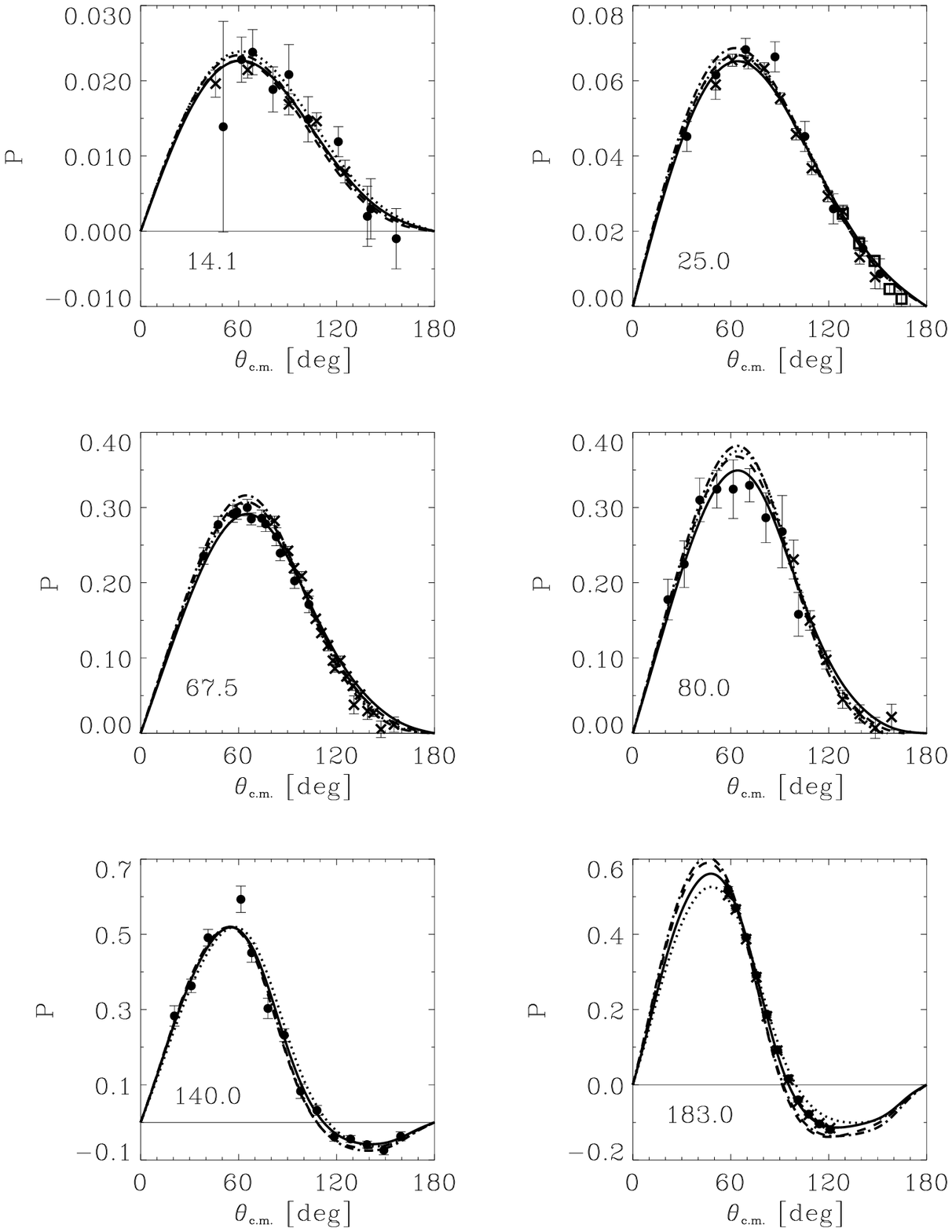,width=16.0cm}
\end{picture}
\caption[Observables of $np$ scattering]
{{\small Observables of $np$ scattering, notations as in
Fig.\,\protect{\ref{npob1}}. 
\label{npob2}}}
\end{figure}

\begin{figure}[t] \centering
\begin{picture}(16.0,20.0)(0.0,2.0)
\epsfig{figure=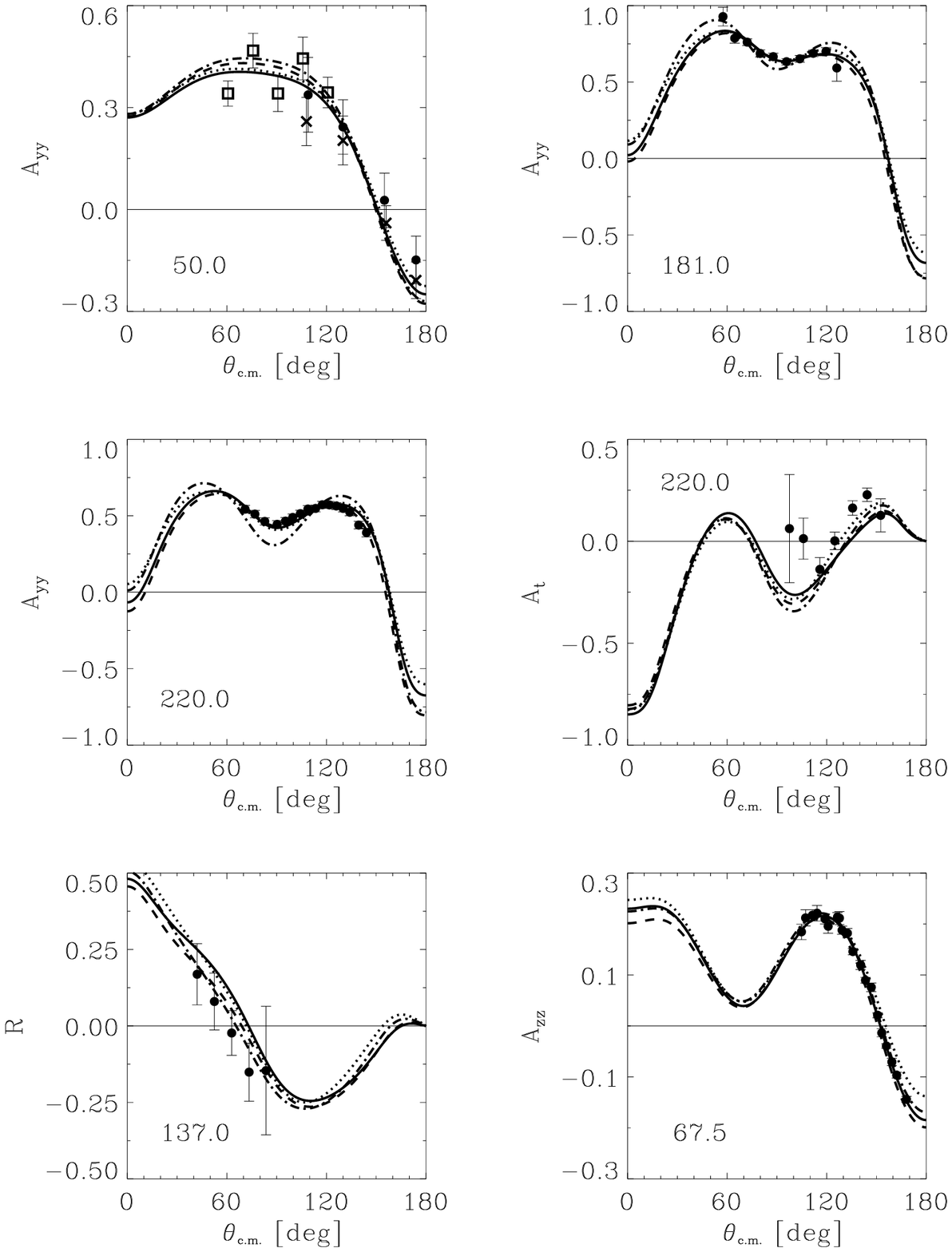,width=16.0cm}
\end{picture}
\caption[Observables of $np$ scattering]
{{\small Observables of $np$ scattering, notations as in
Fig.\,\protect{\ref{npob1}}. 
\label{npob3}}}
\end{figure}

\begin{figure}[t] \centering
\begin{picture}(16.0,20.0)(0.0,2.0)
\epsfig{figure=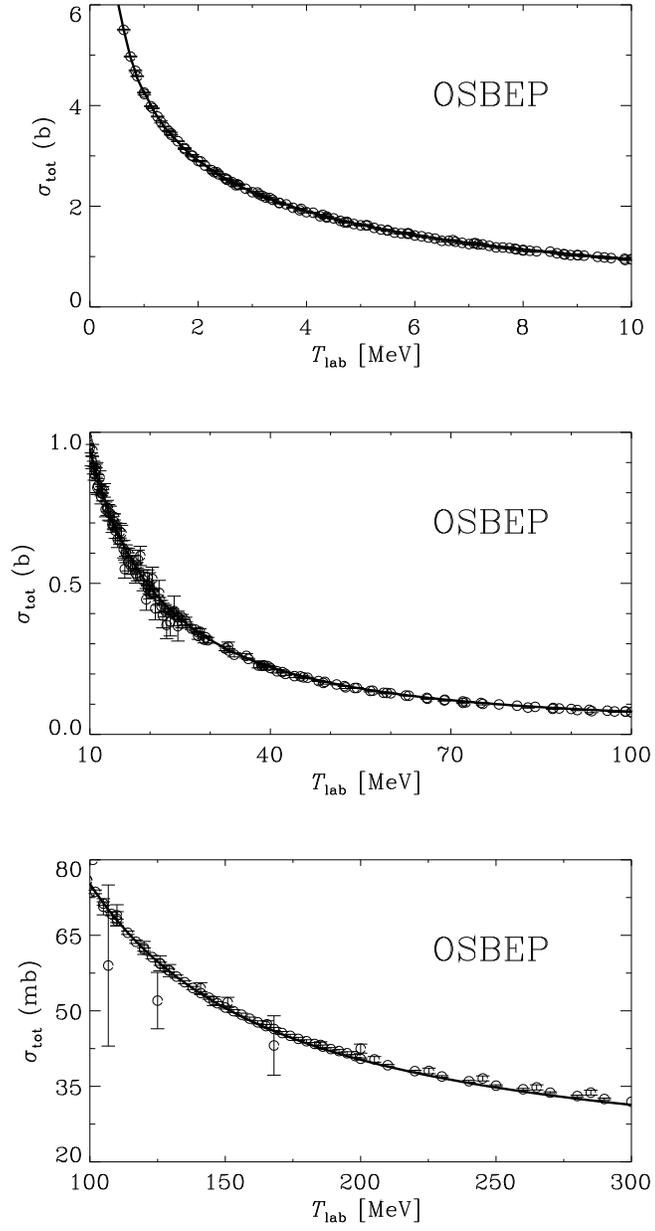,width=16.0cm}
\end{picture}
\caption{Total cross section for elastic $np$ scattering. \label{stot}}
\end{figure}

Altogether, there exist 2719 data points for 13 observables 
between 0 and 300\,MeV. Out of 260 possible plots,  
we selected 19 as representative. They are shown in
Figs.\,\ref{npob1}-\ref{stot}. 
For each measured observable, we plot the
theoretical results of OSBEP (full line), Bonn-B (dashed),
Nijm93 (dotted), and Paris (dash-dotted). 
Visually, the models are hardly to distinguish. 
This is important in view of quite different phase shifts
discussed above and significantly different number of adjustable
parameters. OSBEP uses about half the
parameters of the other models. 
  
However, there are quantitative differences between the models 
as shown for the $\chi^2$/datum listed in Table\,\ref{chisq}. The table 
reflects how the database developed during the
last years. In the meantime, a number of very precise measurements of
differential cross sections and polarization 
observables became available. In
particular, the accurate polarization data at 183\,MeV in 
Fig.\,\ref{npob2} from the IUCF group \cite{said} 
yield large $\chi^2$ contributions for the Nijm93, Paris, and
Bonn-B potential whereas the OSBEP agrees very well with these 
data. Besides
that, the differential cross section measurements in Fig.\,\ref{npob1},
which at large angles are sensitive on the $\pi NN$-coupling constant,
seem to support the low value of 13.75 used in the OSBEP potential (full
line) rather than the older value of 14.4 which is used in the Bonn-B
potential (dashed). Therefore, to have a fair comparison, the 
conventional models should be updated to today's database. 
As yet, the application in $np$ 
scattering shows that OSBEP is able to
describe the data with comparable accuracy as standard $NN$
potentials using eight parameters only which lends support 
for the model of solitary mesons and for the scaling law (\ref{scaling})
in particular.
   
Besides the general excellent agreement in polarization and
spin transfer observables we stress the high
accuracy which is obtained in the description of the
$np$ spin-correlation parameter $A_{zz}$ at 67.5 MeV,
Fig.\,\ref{npob3}, measured by the Basel group
\cite{basel}. In this context, Klomp, Stoks, and
de Swart \cite{KLO92} argue that a potential 
which describes $A_{zz}$ at this energy does not allow a
high $^3SD_1$ mixing angle $\epsilon_1$ at 50 MeV.
Their own PWA, including the Basel data, yields
$\epsilon_1=2.2^{\circ}\pm0.5^{\circ}$ at this energy.
Figure \ref{coupl} shows that all models considered here
predict $\epsilon_1$ slightly below $2^{\circ}$. 
This must be compared to 
$\epsilon_1=2.9^{\circ}\pm0.3^{\circ}$, a value obtained in a 
phase shift analysis based on the Basel data \cite{basel}, and
the SM97 value which is $2.53^{\circ}\pm 0.19^{\circ}$. We are inclined to 
follow the arguments of Klomp {\it et al.} 
that these values are too large. 
A similar argument was given by Machleidt and Slaus \cite{masl}. 
   
The total elastic cross section is 
very well accounted for by OSBEP 
over the whole energy range, see Fig.\,\ref{stot}.
The $\chi^2/\mbox{datum}$ equals 9.5 for 319 data points below 300\,MeV.
This value is surprisingly high and not anticipated from Fig.\,\ref{stot}. 
This is mainly due to one dataset only,
the experiment of Lisowski {\it et al.} \cite{li82}, whose
67 data points are associated with very small error bars, which
contributes a $\chi^2$/datum of 39.
The remaining 252 $\sigma_{tot}$
measurements are fitted with $\chi^2/\mbox{datum}$ of 1.6. 
      
An other quality of fit is obtained by 
the high-precision $NN$ Bonn-CD \cite{bcd} and the Reidlike 
Nijmegen potentials \cite{nipot}. They sacrifice the simplicity of the
original boson exchange potentials and fit each partial wave separately. 
\subsection{$pp$ observables}
\label{ppob}
\begin{figure}[t] \centering
\begin{picture}(16.0,20.0)(0.0,2.0)
\epsfig{figure=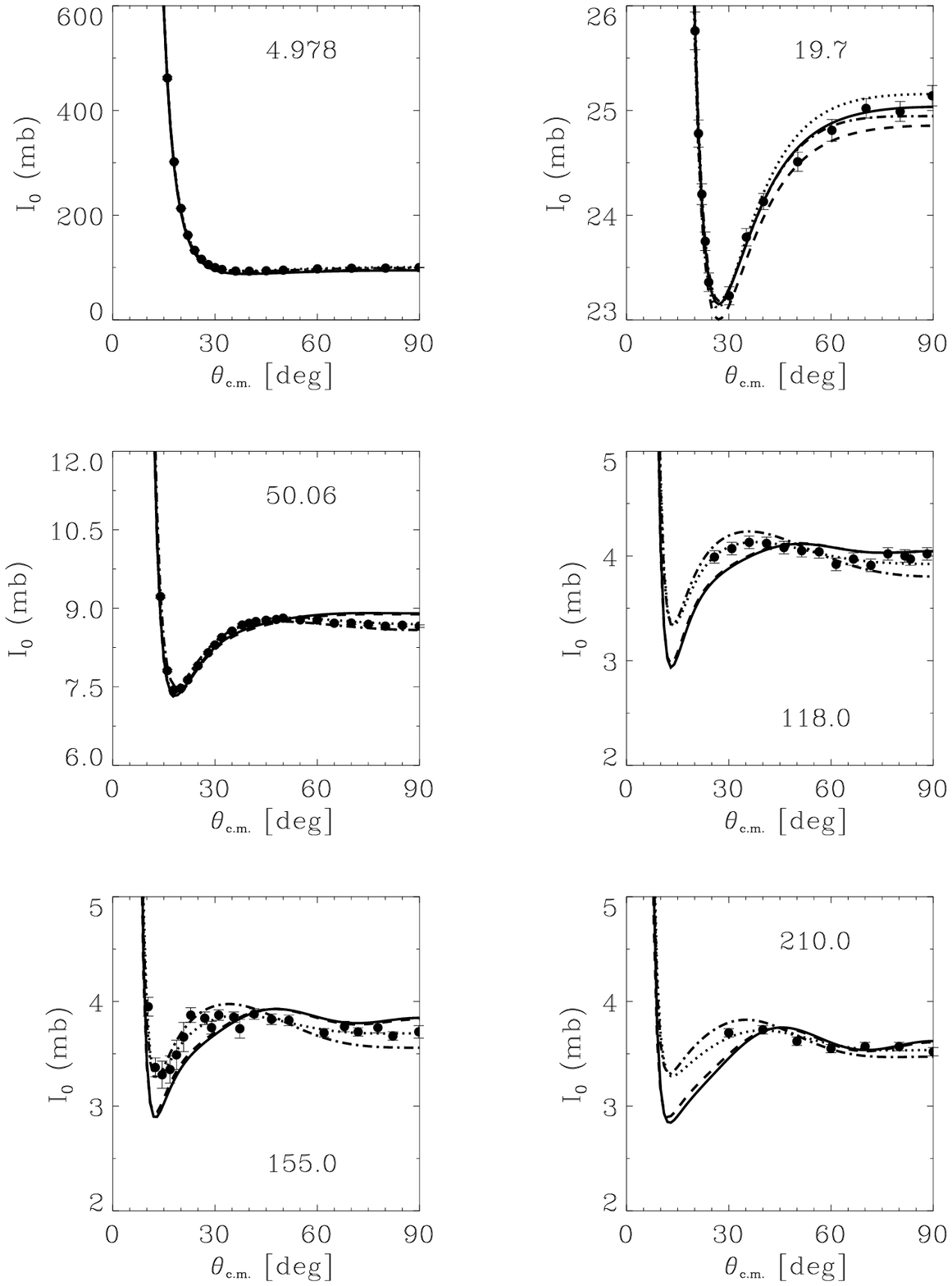,width=16.0cm}
\end{picture}
\caption[Observables of $pp$ scattering]
{Observables of $pp$ scattering. 
Kinetic laboratory energy is denoted, experimental data are taken from
\protect{\cite{said}} with notation from \protect\cite{Hos68}. 
We show theoretical predictions from OSBEP
(full), Bonn-B (dashed, see text), Nijm93 (dotted), and Paris
(dashed-dotted). \label{ppob1}}
\end{figure}

\begin{figure}[t] \centering
\begin{picture}(16.0,20.0)(0.0,2.0)
\epsfig{figure=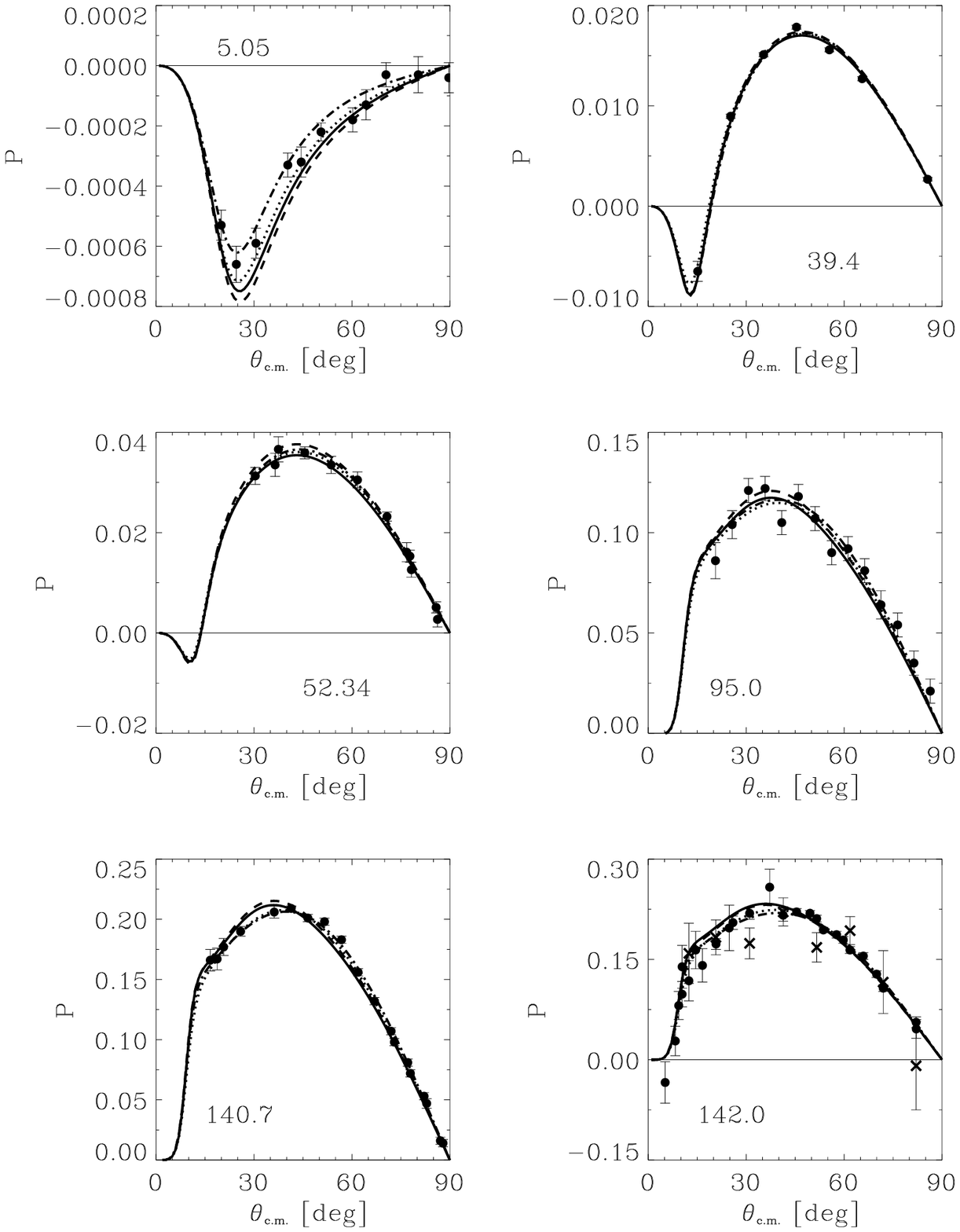,width=16.0cm}
\end{picture}
\caption[Observables of $pp$ scattering]
{{\small Observables of $pp$ scattering, notations as in
Fig.\,\protect{\ref{ppob1}}. 
\label{ppob2}}}
\end{figure}

\begin{figure}[t] \centering
\begin{picture}(16.0,20.0)(0.0,2.0)
\epsfig{figure=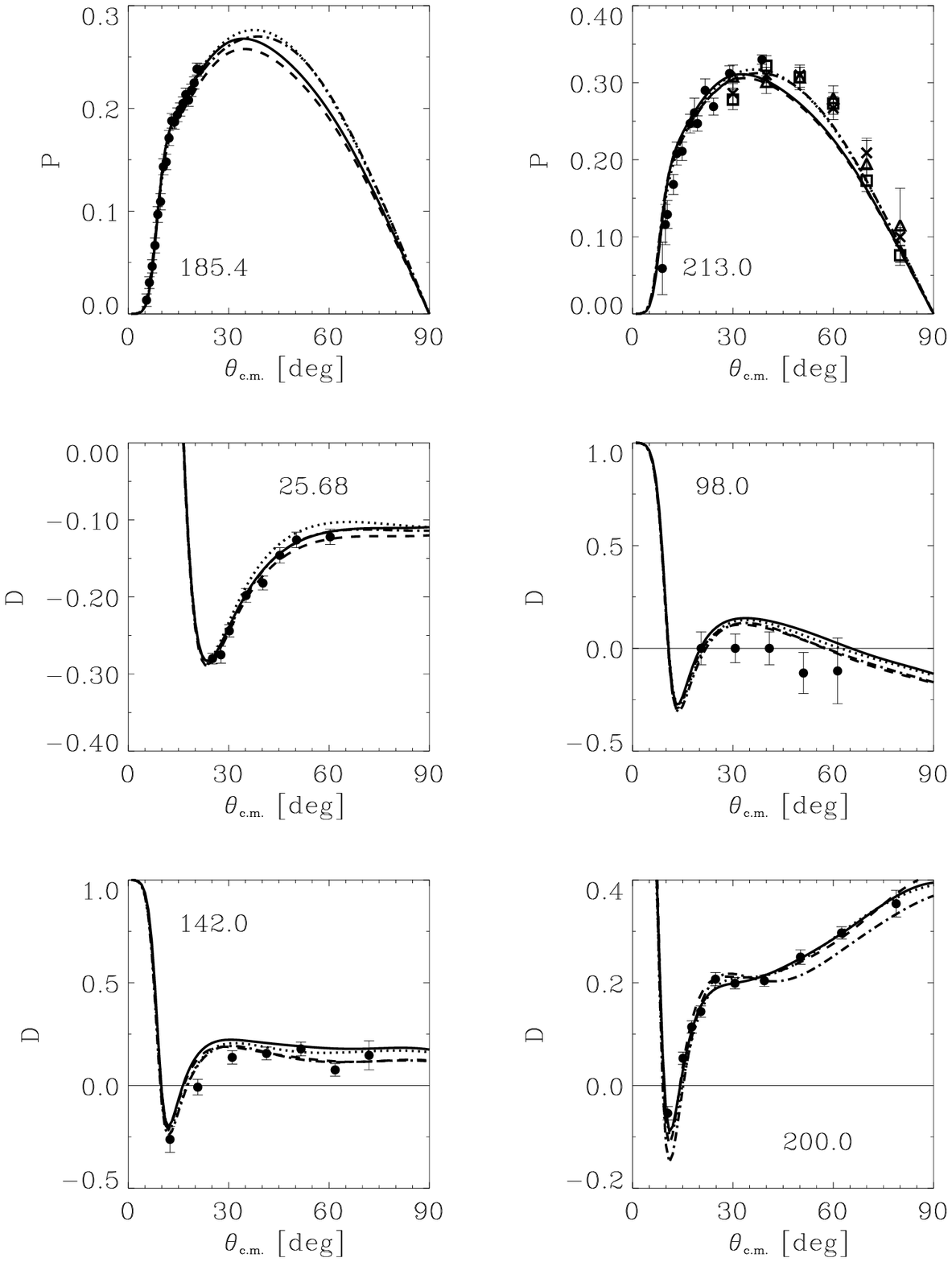,width=16.0cm}
\end{picture}
\caption[Observables of $pp$ scattering]
{{\small Observables of $pp$ scattering, notations as in
Fig.\,\protect{\ref{ppob1}}. 
\label{ppob3}}}
\end{figure}

\begin{figure}[t] \centering
\begin{picture}(16.0,20.0)(0.0,2.0)
\epsfig{figure=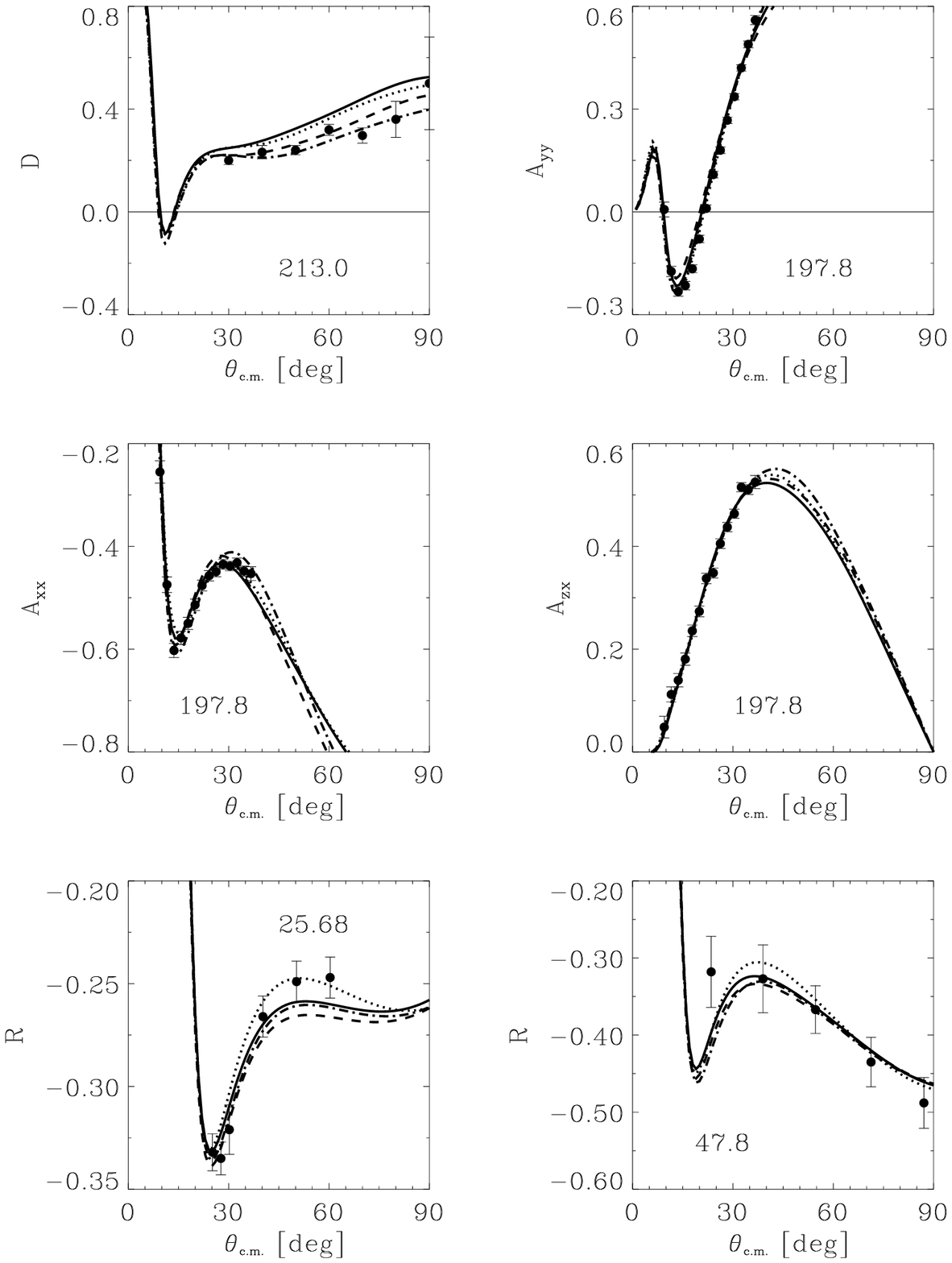,width=16.0cm}
\end{picture}
\caption[Observables of $pp$ scattering]
{{\small Observables of $pp$ scattering, notations as in
Fig.\,\protect{\ref{ppob1}}. 
\label{ppob4}}}
\end{figure}

\begin{figure}[t] \centering
\begin{picture}(16.0,20.0)(0.0,2.0)
\epsfig{figure=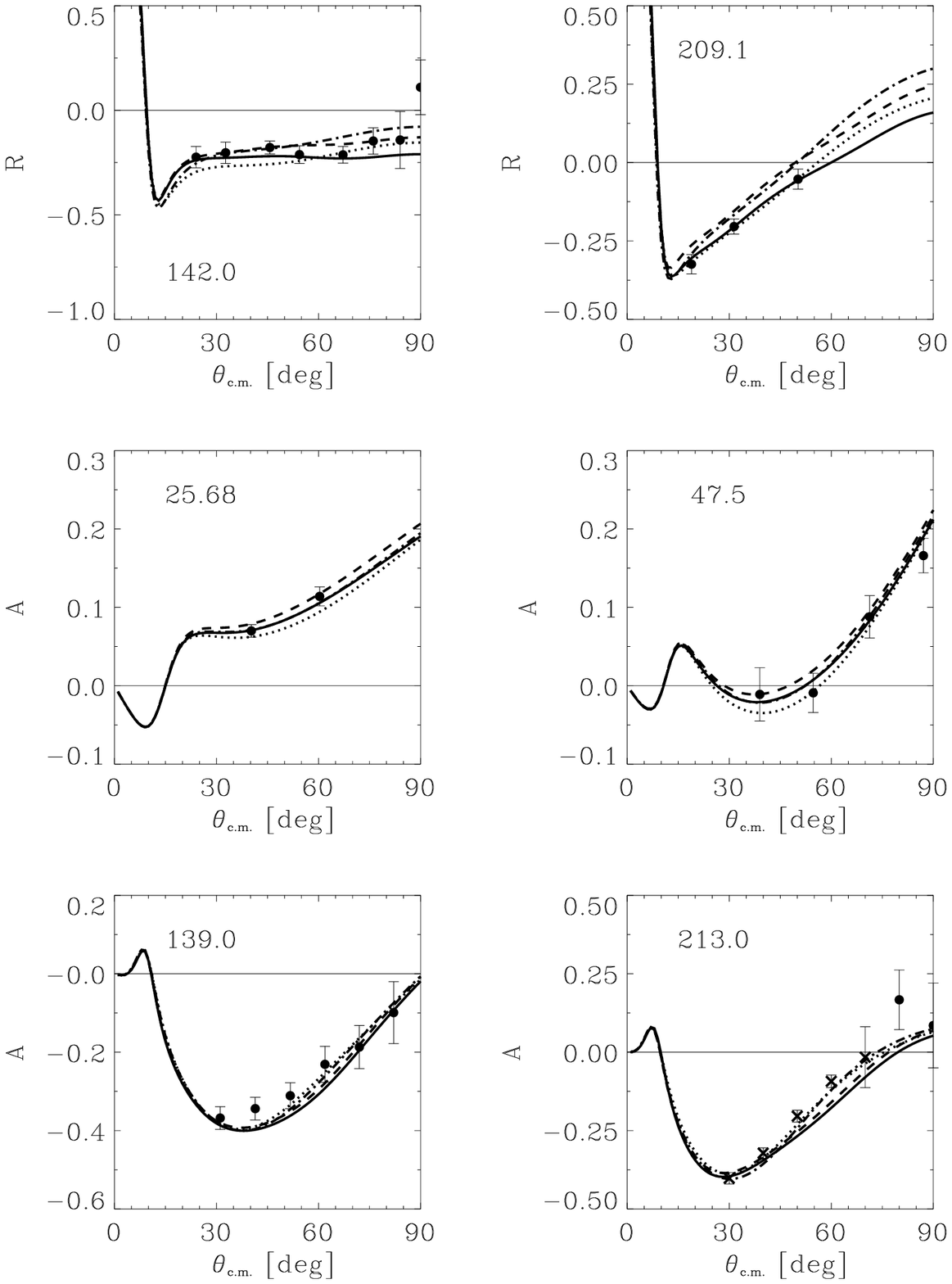,width=16.0cm}
\end{picture}
\caption[Observables of $pp$ scattering]
{{\small Observables of $pp$ scattering, notations as in
Fig.\,\protect{\ref{ppob1}}. 
\label{ppob5}}}
\end{figure}

Experimental data cover the interval 1-300\,MeV.  
The data below 1\,MeV have been discarded 
since an assessment of the low-energy data is difficult in the sense
that the full electromagnetic interaction has to be taken into account,
which is very hard to do in a momentum space calculation. Additionally, 
these data are associated with very small errors and it is misleading to
include them in a $\chi^2$ calculation since they can easily distort the
result \cite{STO}. 
After this subtraction, we are left with
1292 data points for 16 observables. 
In Figs.\,\ref{ppob1}-\ref{ppob5}, we show 30 plots representative for a
total number of 215 possible plots.  
The Paris potential is still a good fit to the $pp$ data. 
As the $\chi^2$ in Table \ref{chisq} indicates, 
the quality of OSBEP is not as good as in the case of
$np$ scattering. This can be traced to the 
overattraction of one boson exchange potentials in $P$ 
channels which signals the lack of $\pi\rho$ correlations in our 
potential. A more detailed discussion can be found in \cite{Mach87}. 
$np$ data are less seriously affected, since isoscalar
contributions partly compensate this shortcoming. 
We have noticed that a significant improvement 
is achieved with an artificially large $\eta NN$ coupling
constant, which contradicts the SU(3) flavor
symmetry constraint (\ref{etann}). 
The full Bonn potential includes $\pi\rho$
correlations and neglects $\eta$ exchange,
putting $g_{\eta}=0$.
The one boson exchange
approximation Bonn-B simulates the same contributions by
using $g_{\eta}^2/{4\pi}=3$. 
We prefer to use a value $g_{\eta}^2/{4\pi}=0.702$ which agrees
with SU(3) symmetry and rely on a more elaborated model,
including $\pi\rho$ contributions and $\Delta$ isobars, to
provide better $P$ wave phase shifts in a future work. 
  
A $pp$ version of Bonn-B does not exist in literature. 
We generate a Bonn-B potential suitable for $pp$ analysis 
by substituting the average nucleon and pion mass of the $np$ version 
by the proton and $\pi^0$ mass, respectively, and including the Coulomb
potential into the scattering equation. In addition, we refitted the
$\sigma_1$ coupling constant to be $g^2_{\sigma_1}/{4\pi}=8.8235$.  
The same prescription was used for OSBEP. 
Since the main contribution to the
large $\chi^2$ comes from the differential cross section
in the energy bin 50-150 MeV, we show some of the
measured cross sections in Fig.\ \ref{ppob1}. 
It is obvious, that OSBEP and Bonn-B yield
almost the same results, in some of the figures the two curves 
cannot be distinguished. We obtain a
value for the Bonn-B $\chi^2$ which is larger than Nijm93
and Paris but slightly below OSBEP. This is consistent with
the enlarged $\eta NN$ coupling constant which somewhat compensates 
the overattraction in the $P$ waves. The remaining harm therefore
sticks with the approximations made concerning 
the meson-meson correlations
whereas the model of solitary bosons and the scaling law find the same
confirmation as deduced from $np$.  
\section{Outlook}
\label{outl}
With this analysis, we make a comparison of $np$ and $pp$ observables
below pion threshold with several potential models. The total
$\chi^2$/datum shows the high standard of all models but also some
consistent failures. For the one boson exchange potentials, they become
obvious for $P$ waves and $pp$ differential cross sections 
above 80\,MeV. This shortcoming is well known from older analyses but is
here confirmed and has its cause in the absence of meson-meson
correlations. The phenomenological form factors have been consistently
replaced by properly normalized solitary meson fields which guarantee
finite self-energies. An empirical scaling law was deduced from
comparison with Bonn-B form factors and this rule was used in case of
$np$ and $pp$ data. This issue permitted the reduction of fit parameters
to the meson-nucleon coupling constants and one parameter accounting
for the meson self-interaction. This study serves the purpose to
consistently describe all $NN$ data below pion threshold with the claim to
be highly quantitative but with significantly reduced degrees of freedom
in the fits. The $\chi^2$/datum results 
are listed in Table\,\ref{chisq}. The OSBEP
result is close to the Nijm93 potential, whereas the Bonn-B
and Paris potential yield considerably larger values. However,
both models may easily be refitted to improve their $\chi^2$
with respect to the latest database. 
   
As our comparison of several potential models and
their predictions for observables of elastic $NN$ scattering shows, 
there is little room for improvements or to 
discern model details on-shell. In previous work \cite{be}, 
we made a strong point that $(p,p\gamma)$ bremsstrahlung, triton binding
energy, and nucleon-nucleus scattering cannot discern off-shell
differences if the on-shell amplitudes are equivalent.
   
The boson exchange models 
cannot be extended towards higher
energies, the regime of meson production, and hadronic
excitations, since this requires a genuine QCD dynamics. 
New experimental facilities, such as IUCF, CELSIUS, COSY, and TJNAF  
provide high-quality data and we are seriously considering various
potential models suitable for this new domain. 
Beyond any doubt, this is a subtle problem. Prior to this, it appears
interesting to investigate the empirical scaling law in more detail and
have a look into the boson exchange model for $\pi N$ scattering.   
In this context, it is a common problem that the form
factor parametrization of the $NN$ interaction can not be
used in the calculation of nucleon pole diagrams
\cite{mnp}. This may be the reason for the failure of 
the attempts to gain a consistent description of $NN$,  
$\pi N$ and $\pi\pi$ interactions. To achieve this goal
would lend support
for the proper normalization of solitary meson fields.  
The higher order diagrams in the $\pi N$
scattering equations, which need to be regularized by a form factor, are
essentially baryon self-energy and vertex correction amplitudes.  
Since the proper normalization was designed to yield finite results for
these diagrams, it is corollary to work also there.
\acknowledgements
The authors appreciate discussions with M. Sander
and would like to thank R.\,A. Arndt 
for providing some {\sc FORTRAN} routines and encoding OSBEP into
{\sc SAID}. One of us (L.\,J.) thanks B. Apagyi for his 
hospitality at the Technical University of Budapest. This work was
supported in part by Forschungszentrum J\"ulich GmbH under Grant No.\
41126865. 
\tighten
%
%
\begin{table} 
\caption{OSBEP parameters.}
\label{coco}
\begin{tabular}{llllllll}
 & $\pi$ & $\eta$ & $\rho$ & $\omega$ & $\sigma_0$ & $\sigma_1$ 
& $\delta$ \\
\hline
$S^P$ & $0^-$ & $0^-$ & $1^-$ & $1^-$ & $0^+$ & $0^+$ & $0^+$ \\
$m_{\beta}$ [MeV]
& 138.03\tablenotemark[1] & 548.8 & 769 & 782.6 & 720 & 550 & 983 \\
 $\displaystyle{\frac{g_{\beta}^2}{4\pi}}$ &
    13.75   & 0.702 & 1.431 & 21.07 & 14.64 & 8.6619\tablenotemark[1]
& 1.259 \\
\multicolumn{4}{c}{$\alpha_{\pi}=0.44065$} &
\multicolumn{4}{c}{$f_{\rho}/g_{\rho}=3.829$} \\
\end{tabular}
\tablenotetext[1]{Values for the $pp$ potential are 
$m_{\pi}=134.9764$\,MeV and $g_{{\sigma_1}}^2/{4\pi}=8.5531$.}
\end{table}
\begin{table} 
\caption{Deuteron properties.}
\label{deut}
\begin{tabular}{lllll}
 & \multicolumn{1}{c}{Bonn-B \protect\cite{Mach89}} &
\multicolumn{1}{c}{OSBEP} & 
\multicolumn{1}{c}{Exp.} &
\multicolumn{1}{c}{Ref.} \\
\hline
$E_B\mbox{ (MeV)}$ & $2.2246$ &    $2.22459$   
&  $2.22458900(22)$ &
\protect\cite{red}  \\   
$ \mu_d$           & $0.8514$\tablenotemark[1]  &  
$0.8524$\tablenotemark[1] &  $0.857406(1)$    & 
\protect\cite{rmu}  \\
$Q_d\mbox{ (fm$^2$)}$ &  $0.2783$\tablenotemark[1] &
$0.2698$\tablenotemark[1] &  $0.2859(3)$      &
\protect\cite{rqa}   \\
$A_S\mbox{ (fm$^{-1/2}$)}$  &  $0.8860$    &
   $0.8805$    &   $0.8802(20)$  &
\protect\cite{rqa}   \\
$D/S$           &   $0.0264$    &           
   $0.0258$    &   $0.0256(4)$   &
\protect\cite{rds}   \\
$r_{RMS}\mbox{ (fm)}$ &  $1.9688$     &      
  $1.957$     &  $1.9627(38)$      &
\protect\cite{rqa}   \\
$P_D\quad(\%)$     &   $4.99$      &          
   $4.80$      &
\multicolumn{1}{c}{ }    
&   \multicolumn{1}{c}{ }  \\
\end{tabular}
\tablenotetext[1]{Meson exchange current contributions not 
included.}    
\end{table}
\begin{table} 
\caption{$\chi^2$/datum for the OSBEP and several potential models. Data
and $\chi^2$ values for the OSBEP, Nijm93 and Paris 
potential were taken from {\sc SAID} \protect\cite{said}.}
\label{chisq}
\begin{tabular}{lllll}
\multicolumn{1}{c}{Model}  & \multicolumn{1}{c}{No.\ of param.} & 
\multicolumn{1}{c}{$np$\tablenotemark[1]}
& \multicolumn{1}{c}{$pp$\tablenotemark[2]} & 
\multicolumn{1}{c}{Total} \\\hline
 OSBEP        & 8  & 4.1 & 6.8 & 5.0 \\
 Nijm93 & 15 & 5.6 & 2.2 & 4.5 \\
 Bonn-B      & 15 &12.1 & 5.8\tablenotemark[3] & 10.1 \\
 Paris        &$\approx$60 & 17.5 & 2.3 & 12.6 \\
\end{tabular}
\tablenotetext[1]{Energy bin 0-300\,MeV (2719 data points).}
\tablenotetext[2]{Energy bin 1-300\,MeV (1292 data points).}
\tablenotetext[3]{$pp$ version
$g_{\sigma_1}^2/{4\pi}=8.8235$, see text.}
\end{table}
\clearpage
%
%
%
\clearpage
%
%
\clearpage
%
%
\clearpage
%
%
\clearpage
%
%
\clearpage
%
%
\clearpage
%
%
\clearpage
%
%
\clearpage
%
%
\clearpage
%
%
\clearpage
%
%
\clearpage
%
%
\end{document}